\documentclass[12pt]{article}

\usepackage{amsfonts,amssymb,amsmath}
\usepackage{graphicx}
\DeclareGraphicsExtensions{epsfig}
\begin{document}
\title{\bf Enhancing quantum state transfer efficiency in binary-tree spin networks by partially collapsing measurements}
\author{ Naghi Behzadi $^{a}$
\thanks{E-mail:n.behzadi@tabrizu.ac.ir}
and Bahram Ahansaz $^{b}$
\\ $^a${\small Research Institute for Fundamental Sciences, University of Tabriz, Tabriz, Iran,}
\\ $^b${\small Physics Department, Azarbaijan Shahid Madani University, Tabriz, Iran}} \maketitle

\begin{abstract}
\noindent
In this work, quantum state transfer (QST) over binary-tree spin networks is studied
by using advantages of partially collapsing measurements. To this aim, we perform initially a weak measurement (WM) on central qubit of the binary-tree network, which encoding the state of concern and after time evolution of the whole system, a quantum measurement reversal (QMR) on the destined qubit is performed. By taking the optimal value of the QMR, it is shown that the QST can be improved considerably by controlling the WM strength and by choosing it close enough to 1, near-perfect QST can be achieved. We also show that how entanglement distribution quality over the binary-tree spin network can be obviously improved by using this approach.
\\
\\
{\bf PACS Nos:}
\end{abstract}

\section{Introduction}
Realizing any advantageous protocol in quantum communication and distributed quantum computation depends on reliable quantum state transfer (QST) from one point to another throughout an efficient quantum communication channel. Obviously, implementing these protocols needs to many body interacting quantum systems such as spin chains \cite{Bose}. In many protocols of QST based on spin chains, the performance of the
protocols depends on the engineering of coupling strengths between the spins such that the perfect
state transfer with uniform coupling is possible only for chains with two and three spins \cite{Christandl}. Therefore, to overcome to these difficulties many efforts, in one hand, have been spent for engineering the coupling strengths in spin chains and, on the other hand, enormous works devoted to external manipulation on the system to achieve perfect or near perfect state transfer [3-20]. It would be worthwhile to have ability to route quantum states from one sender to different recipients in general network structures or graphs, the goal which has attracted several attentions \cite{Lorenzo, Pemberton-Ross, karimipour, Behzadi, Stefanak, Kay, Tsomokos, Bernasconi}. This point, in turn, is an important feature in increasing the connectivity within a quantum computer.

According to the quantum mechanics postulates \cite{Nielsen}, quantum strong measurement of a variable of
a quantum system irrevocably collapses the initial state to one of the eigenstates of the measurement operator, but quantum WM can reveal some information about the amplitudes of a quantum state without collapsing the state into eigenstates. Recently, a scheme has been proposed to improve the QST by applying the WM and QMR on the sending and the receiving qubits, respectively \cite{Man1}. This scheme is applicable not only to the usual QST in a spin chain with uniform coupling strengths, but also to the QST among qubits chain which experience energy dissipations.

In this work, we investigate the QST over a many-body interacting dissipative spin system with underlying binary-tree networks by using the partially collapsing measurements.
Binary-trees are interesting structures which appear in a variety of applications including quantum
algorithms \cite{Childs1,Childs2} and as a possible structure for artificial light-harvesting systems and energy transport \cite{Adronov, Bradshaw, Novo}.
Also, binary-tree configuration is arguably the most significant network topology in circuit design.
Therefore, implementation of the QST protocol over this configurations is an essential task.
We show that how the combination of QST scheme with partially collapsing measurements on the binary tree networks
leads to the significant improvement in the fidelity of received state.

On the other hand, it was pointed out that WM and QMR can effectively protect the quantum states of a single qubit system from decoherence \cite{Korotkov, Lee}. Also it is shown that the WM and its reversal counterpart can greatly protect the entanglement of two-qubit systems from amplitude damping
decoherence \cite{Sun, Kwon}. Also, the robust state transfer together with entanglement distribution in a linear spin chain using the WM and the respective reversal one have been studied in \cite{He}.
In this regards, it would be worthwhile to investigate the effect of WM and QMR on the entanglement distribution quality over the binary-tree network in this paper. It is found out that the entanglement distribution efficiency can be obviously improved by using this approach.

The paper is structured as follows: In Sec. II, we introduce the binary-tree spin network where each of the qubits interacts with a dissipative environment independently. In Sec. III, we demonstrate the effect of WM on the sending qubit and obtain the dynamical evolution of the system exactly. Sec. IV is devoted to explain the QMR and the reconstruction of the transferred state. In Sec. V, we extend the advantages of the previous sections on providing the entanglement distribution over the network. Finally, we give a brief conclusion in Sec. VI.

\section{The model}
We start by considering QST in a network of interacting qubits with underlying binary-tree graph as one that depicted in Fig. 1. Each vertex is corresponding to a qubit with transition frequency $\omega_{0}$, and the edges represent
the corresponding couplings with strength $\nu$. The Hamiltonian for this interacting system is considered as
\begin{eqnarray}
H_{tree}=\omega_{0}\sum_{j=1}^{2^{N}-1}|j\rangle\langle j|+\nu\sum_{j=1}^{2^{N-1}-1}(|j\rangle\langle 2j|+|j\rangle\langle2j+1|+h.c.),
\end{eqnarray}
where $|j\rangle=|0\rangle^{\otimes j-1}\otimes|1\rangle\otimes|0\rangle^{\otimes2^{N}-j-1}$, and so $\{|j\rangle\}_{j=1}^{2^{N}-1}$ is corresponding to the standard set of basis for the single excitation subspace for the Hilbert space of the tree-network with $N$ generations. Suppose that the state of concern is
\begin{eqnarray}
|\psi(0)\rangle=cos(\theta/2)|0\rangle+e^{i\phi}sin(\theta/2)|1\rangle,
\end{eqnarray}
which is prepared on the $j=1$ spin. The main goal is that, it is enabled by the partially collapsing measurements to transfer the state (2) from the site 1 to any other site in the network with a reliable fidelity. Each spin of the network interacts with an dissipative environment independently and therefore, the interaction Hamiltonian can be written as
\begin{eqnarray}
H_{int}=\sum_{j=1}^{2^{N}-1}\sum_{k}(g_{k}|j\rangle\langle 0|b_{k}^{j}+g_{k}^{\ast}|0\rangle\langle j|b_{k}^{j\dagger}).
\end{eqnarray}
where $g_{k}$ is the strength of coupling of the spin on site $j$ to the $k$th mode of the environment and $b_{k}^{j}$ $(b_{k}^{j\dagger})$ is the corresponding annihilation (creation) operator. Let's consider an unitary transformation as follow
\begin{eqnarray}
U=\bigoplus_{m=0}^{N}H^{\otimes m},
\end{eqnarray}
where $H$ is the well-known single qubit Haddamard matrix (as an example for $N=4$, $U=1\oplus H\oplus H\otimes H\oplus H\otimes H \otimes H\oplus H\otimes H\otimes H\otimes H$). It is easy to show that under this unitary transformation, the Hamiltonian (1) takes a block diagonal form where the corresponding largest block is as
\begin{eqnarray}
H_{tree}^{C}=\omega_{0}\sum_{m=1}^{N}|C_{m}\rangle\langle C_{m}|+\sqrt{2}\nu\sum_{m=1}^{N-1}(|C_{m}\rangle\langle C_{m+1}|+|C_{m+1}\rangle\langle C_{m}|),
\end{eqnarray}
and the corresponding invariant subspace is
\begin{eqnarray}
\mathcal{H}_{tree}^{C}=span\{|C_{m}\rangle\},\\\nonumber&&\hspace{-73mm}
\qquad|C_{m}\rangle\equiv\frac{1}{\sqrt{2^{m-1}}}\sum_{j=2^{m-1}}^{2^{m}-1}|j\rangle, \qquad m=1, 2, ..., N.
\end{eqnarray}
Obviously, the state $|C_{m}\rangle$ is corresponding to the $m$th generation of the tree whose initial state is $|1\rangle$. As is clear, the state $|1\rangle$ belongs to the this subspace so, if at the initial time the state of the concern in (2) prepared at the sit 1, the evolved state belongs to it too. Also, the Hamiltonian (5) is similar to the Hamiltonian of spin chain with $N$ site, and with uniform nearest neighbor interaction of strength $\sqrt{2}\nu$ and therefore, it is exactly solvable. So in this way, the interaction Hamiltonian corresponding to the Hamiltonian (5) becomes as
\begin{eqnarray}
H_{int}^{C}=\sum_{m=1}^{N}\sum_{k}\left(g_{k}|C_{m}\rangle\langle 0|B_{k}^{m}+g_{k}^{\ast}|0\rangle\langle C_{m}|B_{k}^{m\dagger}\right),
\end{eqnarray}
where
\begin{eqnarray}
B_{k}^{m}=\frac{1}{\sqrt{2^{m-1}}}\sum_{j=2^{m-1}}^{2^{m}-1}b_{k}^{j}.
\end{eqnarray}
In the interaction picture, the Hamiltonian corresponding to the largest invariant subspace becomes as
\begin{eqnarray}
H_{int}^{C}=\sqrt{2}\nu\sum_{m=1}^{N-1}(|C_{m}\rangle\langle C_{m+1}|+|C_{m+1}\rangle\langle C_{m}|)\\\nonumber&&\hspace{-98mm}+\sum_{m=1}^{N}\sum_{k}\left(g_{k}^{m}|C_{m}\rangle\langle 0|B_{k}^{m}(t)e^{i\omega_{0}t}+g_{k}^{m\ast}|0\rangle\langle C_{m}|B_{k}^{m\dagger}(t)e^{-i\omega_{0}t}\right),
\end{eqnarray}
where
\begin{eqnarray}
B_{k}^{m}=\frac{1}{\sqrt{2^{m-1}}}\sum_{j=2^{m-1}}^{2^{m}-1}b_{k}^{j}e^{-i\omega_{k}^{j}t}.
\end{eqnarray}

\section{WM and dynamical evolution of the system}

We now consider QST on the binary-tree network sketched in Fig. 1. We show that the combination of QST scheme with partially collapsing measurements, namely a WM followed by a QMR, leads to the significant improvement of the fidelity of the scheme. Let's suppose that at initial time the state (2) is prepared in site 1, and the transferring it to a typical r$th$ site on the network with a reliable fidelity is our demand. At the first step, before the evolution of the system, a WM is performed on the 1$th$ qubit
with strength p with explicit form as
\begin{eqnarray}
\mathcal{W}(p)=\left(
                 \begin{array}{cc}
                  \sqrt{1-p} & 0 \\
                   0 & 1 \\
                 \end{array}
               \right).
\end{eqnarray}
After this measurement, the state (2), after normalization, becomes
\begin{eqnarray}
|\psi(0,p)\rangle=\frac{1}{\sqrt{\mathcal{P}(0,p)}}\biggr(\mathrm{cos}(\theta/2)|0\rangle+e^{i\phi}\mathrm{sin}(\theta/2)\sqrt{1-p}|C_{1}\rangle\biggr),
\end{eqnarray}
where $\mathcal{P}(0,p)=\mathrm{cos}^{2}(\theta/2)+\mathrm{sin}^{2}(\theta/2)(1-p)$ being the success probability without completely
collapsing the measured state via the WM. Obviously, for the Hamiltonian (1) as a closed system, the total number of excitation is conserved. In the same way, if (1) considered as an open system, the total number of excitation of the system and the related environments is also preserved. Let's assume that, at $t=0$, the state of the whole system is
\begin{eqnarray}
|\Psi(0,p)\rangle=\frac{1}{\sqrt{\mathcal{P}(0,p)}}\biggr(\mathrm{cos}(\theta/2)|0\rangle|0\rangle+e^{i\phi}\mathrm{sin}(\theta/2)\sqrt{1-p}|C_{1}\rangle|0\rangle\biggr),
\end{eqnarray}
where the environments are in the their respective vacuum states. By noting that the dynamics of the state $|0\rangle|0\rangle$ is trivial and keeps invariant in time, therefore the time evolution of the state (13) at $t>0$ reads
\begin{eqnarray}
|\Psi(t,p)\rangle=\frac{1}{\sqrt{\mathcal{P}(t,p)}}\biggr(cos(\theta/2)|0\rangle|0\rangle
\\\nonumber&&\hspace{-70mm}+e^{i\phi}sin(\theta/2)\sqrt{1-p}\Bigr(C_{n}(t)|C_{n}\rangle|0\rangle+\sum_{m(\neq n)=1}^{N}C_{m}(t)|C_{m}\rangle|0\rangle+
\sum_{j=1}^{2^{N}-1}\sum_{k}c_{k}^{j}(t)|0\rangle|1_{k}\rangle_{j}\Bigr)\biggr),
\end{eqnarray}
where the state $|r\rangle$ is contained in the state $|C_{n}\rangle$ corresponding to the $n$th generation of the tree network and $|1_{k}\rangle_{j}$ indicates that the $k$th mode of the $j$th reservoir (the reservoir which interacts with the qubit at the j$th$ site of the network). It is supposed that at initial time, $c_{k}^{j}(0)=0$. The time dependent coefficients $C_{m}(t)$ and $c_{k}^{m}(t)$ with $m=1, 2, ..., N$, are determined by the solving following Schrodinger equation in the interaction picture
\begin{eqnarray}
i\frac{d}{dt}|\Psi(t,p)\rangle=H_{I}^{L}(t)|\Psi(t,p)\rangle.
\end{eqnarray}
Eq. 15, leads to the following set of equations
\begin{eqnarray}
i\dot{C}_{m}(t)=\sqrt{2}\nu(C_{m-1}(t)+C_{m+1}(t))+\frac{1}{\sqrt{2^{m-1}}}\sum_{j=2^{m-1}}^{2^{m}-1}\sum_{k}g_{k}c_{k}^{j}(t)e^{i(\omega_{0}-\omega_{k}^{j})t},
\end{eqnarray}
and
\begin{eqnarray}
i\dot{c}_{k}^{j}(t)=\frac{1}{\sqrt{2^{m-1}}}g_{k}^{\ast}C_{m}(t)e^{-i(\omega_{0}-\omega_{k}^{j})t},
\end{eqnarray}
where $m=1, 2, ..., N$, $j=2^{m-1}, ..., 2^{m}-1$ and the conventions that $C_{0}(t)=C_{N+1}(t)=0$.
Integrating Eqs. (17) with initial conditions $c_{k}^{j}(t)=0$ and inserting their solutions into Eq. 16, yield a closed set of integro-differential equations for $C_{m}(t)$s as
\begin{eqnarray}
\dot{C}_{m}(t)=-i\sqrt{2}\nu(C_{m-1}(t)+C_{m+1}(t))-\frac{1}{2^{m-1}}\sum_{j=2^{m-1}}^{2^{m}-1}\int_{0}^{t}dt'\sum_{k}|g_{k}|^{2}e^{i(\omega_{0}-\omega_{k}^{j})(t-t')}C_{m}(t').
\end{eqnarray}
In the right hand side of Eq. 18, the kernel $f_{j}(t-t')=\sum_{k}|g_{k}|^{2}(t)e^{i(\omega_{0}-\omega_{k}^{j})(t-t')}$ is the $j$th reservoir-correlation function which in the limit of large number of reservoir modes, it can be well
approximated by the integration as $f(t-t')=\int d\omega S(\omega)e^{i(\omega_{0}-\omega)(t-t')}$. Since all of the $2^{N}-1$ two-level systems are identical and so are their respective reservoirs, we have dropped the index $j$. $S(\omega)$ is the effective spectral density for a typical reservoir assumed to be Lorentzian as
\begin{eqnarray}
S(\omega)=\frac{1}{2\pi}\frac{\gamma\lambda^{2}}{(\omega-\omega_{0})^{2}+\lambda^{2}},
\end{eqnarray}
where the parameter $\lambda$ indicates the spectral width of the coupling and the parameter $\gamma$ is the coupling constant.
By this consideration, Eq. 18 becomes as follows
\begin{eqnarray}
\dot{C}_{m}(t)=-i\sqrt{2}\nu(C_{m-1}(t)+C_{m+1}(t))-\int_{0}^{t}dt'f(t-t')C_{m}(t').
\end{eqnarray}
By means of a Laplace transformation, Eq. (20) becomes as
\begin{eqnarray}
p\tilde{C}_{m}(p)-C_{m}(0)=-i\sqrt{2}\nu(\tilde{C}_{m-1}(p)+\tilde{C}_{m+1}(p))-\frac{\gamma\lambda}{2(p+\lambda)}\tilde{C}_{m}(p),
\end{eqnarray}
with initial conditions $C_{1}(0)=1$, and $C_{m}(0)=0$ ($m=2, 3, ... N$). To obtain the explicit form for $\tilde{C}_{m}(p)$s, we assume that they connect to the new $\tilde{C'}_{m}(p)$s by the following unitary transformation as
\begin{eqnarray}
\tilde{C}_{m}(p)=\sqrt{\frac{2}{N+1}}\sum_{l=1}^{N}sin(\frac{m l}{N+1}\pi)\tilde{C'}_{l}(p).
\end{eqnarray}
By this transformation, $\tilde{C'}_{m}(p)$s are calculated easily from Eqs. (21) as follows
\begin{eqnarray}
\tilde{C'}_{m}(p)=\sqrt{\frac{2}{N+1}}\frac{(p+\lambda)sin(\frac{m\pi}{N+1})}{p^{2}+p\biggr(\lambda-2i\sqrt{2}\nu cos(\frac{N-m+1}{N+1}\pi)\biggr)+\lambda\biggr(\gamma-4i\sqrt{2}\nu cos(\frac{N-m+1}{N+1}\pi)\biggr)/2},
\end{eqnarray}
with the poles
\begin{eqnarray}
p_{1}=\frac{-\lambda+2i\sqrt{2}\nu cos(\frac{N-m+1}{N+1}\pi)+\sqrt{\Delta_{m}}}{2}, \quad p_{2}=\frac{-\lambda+2i\sqrt{2}\nu cos(\frac{N-m+1}{N+1}\pi)-\sqrt{\Delta_{m}}}{2}.
\end{eqnarray}
where $\Delta_{m}=\bigr(\lambda+2i\sqrt{2}Jcos(\frac{N-m+1}{N+1}\pi)\bigr)^{2}-2\gamma\lambda$.
After inverse Laplace transformation characterized as
\begin{eqnarray}
C'_{m}(t)=\lim_{p\rightarrow p_{k}}\sum_{k=1}^{2}(p-p_{k})\tilde{C'}_{m}(p)e^{p_{k}t},
\end{eqnarray}
Then
\begin{eqnarray}
C'_{m}(t)=\sqrt{\frac{2}{N+1}}\mathrm{e}^{-\bigr(\lambda-2i\sqrt{2}\nu \mathrm{cos}(\frac{N-m+1}{N+1}\pi)\bigr)t/2}\biggr(\mathrm{cosh}(\sqrt{\Delta_{m}}t/2)
\\\nonumber&&\hspace{-95mm}+\frac{\lambda+2i\sqrt{2}\nu \mathrm{cos}(\frac{N-m+1}{N+1}\pi)}{\sqrt{\Delta_{m}}}\mathrm{sinh}(\sqrt{\Delta_{m}}t/2)\biggr)\mathrm{sin}(\frac{m\pi}{N+1}).
\end{eqnarray}
Consequently, by taking inverse Laplace transform from Eqs. (22), the explicit form of $C_{m}(t)$ is obtained.

\section{QMR and reconstruction of the transferred state}
At this step, for an arbitrary time, to retrieve the state of concern (2) at the r$th$ qubit contained in the n$th$ generation of the binary-tree network, performing a QMR with strength $q$ on this qubit transforms the state (14) into the following state
\begin{eqnarray}
|\Psi(t,p,q)\rangle=\frac{1}{\sqrt{\mathcal{P}(t,p,q)}}\biggr(\mathrm{cos}(\theta/2)\sqrt{1-q}|0\rangle|0\rangle+e^{i\phi} \mathrm{sin}(\theta/2)\sqrt{1-p}\frac{C_{n}(t)}{\sqrt{2^{n-1}}}|r\rangle|0\rangle
\\\nonumber&&\hspace{-130mm}+e^{i\phi} \mathrm{sin}(\theta/2)\sqrt{(1-p)(1-q)}\Bigr(\frac{C_{n}(t)}{\sqrt{2^{n-1}}}\sum_{j(\neq r)=2^{n-1}}^{2^{n}-1}|j\rangle|0\rangle+\sum_{m(\neq n)=1}^{N}C_{m}(t)|C_{m}\rangle|0\rangle
\\\nonumber&&\hspace{-130mm}+\sum_{j=1}^{2^{N}-1}\sum_{k}c_{k}^{j}(t)|0\rangle|1_{k}\rangle_{j}\Bigr)\biggr),
\end{eqnarray}
where $\mathcal{P}(t,p,q)=(1-q)\mathrm{cos}^{2}(\theta/2)+(1-p)\mathrm{sin}^{2}(\theta/2)\frac{|C_{n}(t)|^{2}}{2^{n-1}}+(1-p)(1-q)\mathrm{sin}^{2}(\theta/2)(1-\frac{|C_{n}(t)|^{2}}{2^{n-1}})$ is the success probability of the two measurements without completely collapsing the system's state and $2^{n-1}\leq r\leq 2^{n}-1$. To recover the transferred state at the r$th$ qubit, the strength
of the post QMR, i.e. $q$, is judiciously determined in term of the prior WM with strength $p$ and the evolution time $t$. To this aim, we assume $q=1-(1-p)|C_{n}(t)|^{2}/2^{n-1}$ and $C_{n}(t)=|C_{n}(t)|e^{i\phi_{n}(t)}$, so Eq. (27) becomes as follows
\begin{eqnarray}
|\Psi(t,p)\rangle=\frac{1}{\sqrt{\mathcal{P}(t,p)}}\sqrt{1-p}\frac{|C_{n}(t)|}{\sqrt{2^{n-1}}} \biggr(\Bigr(\mathrm{cos}(\theta/2)|0\rangle+e^{i(\phi+\phi_{n}(t))} \mathrm{sin}(\theta/2)|r\rangle\Bigr)|0\rangle
\\\nonumber&&\hspace{-130mm}+e^{i\phi} \mathrm{sin}(\theta/2)\sqrt{1-p}\Bigr(\frac{C_{n}(t)}{\sqrt{2^{n-1}}}\sum_{j(\neq r)=2^{n-1}}^{2^{n}-1}|j\rangle|0\rangle+\sum_{m(\neq n)=1}^{N}C_{m}(t)|C_{m}\rangle|0\rangle
\\\nonumber&&\hspace{-130mm}+\sum_{j=1}^{2^{N}-1}\sum_{k}c_{k}^{j}(t)|0\rangle|1_{k}\rangle_{j}\Bigr)\biggr),
\end{eqnarray}
where $\mathcal{P}(t,p)=(1-p)\frac{|C_{n}(t)|^{2}}{2^{n-1}}\bigr(1+(1-p)sin^{2}(\theta/2)(1-\frac{|C_{n}(t)|^{2}}{2^{n-1}})\bigr)$ is the optimal success probability. In order to have the state
of the r$th$ qubit coincide with that of the state (2), a phase shift realized by the unitary operator
$\mathcal{U}(\phi_{n}(t)) =\{\{e^{-i\phi_{n}(t)}, 0\}, \{0, 1\}\}$ should be generated for the r$th$ qubit, so the state (28) reads
\begin{eqnarray}
|\Phi(t,p)\rangle=\frac{1}{\sqrt{\mathcal{P}(t,p)}}\sqrt{1-p}\frac{|C_{n}(t)|}{\sqrt{2^{n-1}}} \biggr(\Bigr(\mathrm{cos}(\theta/2)|0\rangle+e^{i\phi} \mathrm{sin}(\theta/2)|r\rangle\Bigr)|0\rangle
\\\nonumber&&\hspace{-130mm}+e^{i\phi} \mathrm{sin}(\theta/2)\sqrt{1-p}\Bigr(\frac{C_{n}(t)}{\sqrt{2^{n-1}}}\sum_{j(\neq r)=2^{n-1}}^{2^{n}-1}|j\rangle|0\rangle+\sum_{m(\neq n)=1}^{N}C_{m}(t)|C_{m}\rangle|0\rangle
\\\nonumber&&\hspace{-130mm}+\sum_{j=1}^{2^{N}-1}\sum_{k}c_{k}^{j}(t)|0\rangle|1_{k}\rangle_{j}\Bigr)\biggr).
\end{eqnarray}
To obtain the closeness of the transferred state to the initial state, we use the fidelity which is defined as
$F(t,p)=\langle \psi(0,p)|\rho_{r}(t,p)|\psi(0,p)\rangle$, where $\rho_{r}(t,p)$ is the actual reduced density matrix of the r$th$ qubit. It is clear that the fidelity turns out state-dependent while this scheme is valid for any unknown state, so by averaging $F(t,p)$ over all the pure states on the Bloch sphere we obtain the averaged fidelity as
\begin{eqnarray}
F_{ave}(t,p)=\frac{1}{2}+\frac{1}{(1-p)(1-\frac{|C_{n}(t)|^{2}}{2^{n-1}})}-\frac{\mathrm{ln}(1+(1-p)(1-\frac{|C_{n}(t)|^{2}}{2^{n-1}}))}{(1-p)^{2}(1-\frac{|C_{n}(t)|^{2}}{2^{n-1}})^{2}}.
\end{eqnarray}
In the case of natural evolution of the system, the averaged fidelity takes the form
\begin{eqnarray}
F_{ave}(t,0)=\frac{1}{2}+\frac{1}{3} \frac{|C_{n}(t)|}{\sqrt{2^{n-1}}}+\frac{1}{6} \frac{|C_{n}(t)|^{2}}{2^{n-1}}.
\end{eqnarray}
Moreover, the averaged success probability for the state transfer can be obtained as
\begin{eqnarray}
\mathcal{P}^{^{ST}}_{ave}(t,p)=\frac{1}{2}(1-p) \frac{|C_{n}(t)|^{2}}{2^{n-1}} (2+(1-p)(1-\frac{|C_{n}(t)|^{2}}{2^{n-1}})).
\end{eqnarray}
In Fig. 2, the time-dependence of the averaged fidelity and the corresponding averaged success probability have been shown on the WM strength $p$
for our binary-tree spin network with $N = 4$ and $N = 8$ generations.
It is observed that the fidelities under measurement-controlled evolution, i.e. $F_{ave}(t,p)$, are manifestly larger than the fidelities under natural evolution, i.e. $F_{ave}(t,0)$,
where $F_{ave}(t,0)$ will eventually decay to 0.5.
Interestingly it is shown that, the fidelity after some oscillatory behavior in time reaches the steady value even after a very long time limit.
Also, it should be noted that the averaged fidelity is approaching 1 when $p$ is tending to 1.
By comparing the averaged fidelities and the corresponding averaged success probability, we can see that
by increasing $p$, $F_{ave}(t,p)$ is increasing while $\mathcal{P}^{^{ST}}_{ave}(t,p)$ is becoming vanishingly small.
Moreover, it is obvious that the averaged fidelities with $N = 8$ generations oscillate with less fluctuations amplitude in comparison to the case with $N = 4$ generation,
implying that by increasing the rank of the generation of the binary-tree spin networks, the fidelities reach quickly to their respective steady values.

\section{Entanglement distribution}
In this section, we study entanglement distribution over binary-tree spin networks by using the previously discussed approach.
We assume a general bipartite entangled state in the form
\begin{eqnarray}
|\varphi(0,0)\rangle=cos(\theta/2) |0,1\rangle+e^{i\phi}sin(\theta/2) |1,0\rangle,
\end{eqnarray}
which has been shared between the qubit 1 of the network and a non-interacting qubit represented a qubit 0, as depicted in Fig. 1. Our aim is to show that how the non-interacting 0$th$ qubit and r$th$ qubit of the networks will establish entanglement by the partially collapsing measurements. Similar to the previous discussion, before the evolution of the system we implement the WM with strength $p$ on the 1$th$ qubit, and after the evolution at time $t$, wo do QMR with strength $q$ on the r$th$ qubit. The corresponding reduced density matrix between the 0$th$ and r$th$ qubits, in the standard computational basis $\{|0,0\rangle, |0,1\rangle, |1,0\rangle, |1,1\rangle\}$, becomes
\begin{eqnarray}
\rho_{0,r}(t,p,q)=\left(
\begin{array}{cccc}
  \rho_{11}(t) & 0 & 0 & 0 \\
  0 & \rho_{22}(t) & \rho_{23}(t) & 0 \\
  0 & \rho_{32}(t) & \rho_{33}(t) & 0 \\
  0 & 0 & 0 & 0
  \end{array}
   \right),
\end{eqnarray}
where the elements are
\begin{eqnarray}
\begin{array}{c}
\rho_{11}(t)=\mathrm{cos}^{2}(\theta/2)(1-p)(1-q)/\mathcal{P}^{ED}(t,p,q),\\\\
\rho_{22}(t)=\mathrm{cos}^{2}(\theta/2)q(1-p)|f|^2/\mathcal{P}^{ED}(t,p,q),\\\\
\rho_{33}(t)=\mathrm{sin}^{2}(\theta/2)(1-q)/\mathcal{P}^{ED}(t,p,q),\\\\
\rho_{23}(t)=\rho_{32}^{*}(t)=e^{i\phi} \mathrm{cos}(\theta/2) \mathrm{sin}(\theta/2)\sqrt{1-p}\sqrt{1-q}f/\mathcal{P}^{ED}(t,p,q).
\end{array}
\end{eqnarray}
where $f=C_{n}(t)/\sqrt{2^{n-1}}$ and $\mathcal{P}^{ED}(t,p,q)$ is the success probability of the WM and QMR for entanglement distribution, whose explicit form is
\begin{eqnarray}
\mathcal{P}^{ED}(t,p,q)=\bigr(1-p \mathrm{cos}^{2}(\theta/2)\bigr)(1-q)+\mathrm{cos}^{2}(\theta/2) q(1-p)|f|^2.
\end{eqnarray}

To quantify the amount of entanglement of the state $\rho_{0,r}(t,p,q)$, we use concurrence as a measure of two-qubit entanglement $\cite{Wootterrs}$.
The following analytical closed form for the concurrence is obtained
\begin{eqnarray}
C\bigr(\rho_{0,r}(t,p,q)\bigr)=2 \mathrm{max}\{0,\mathrm{cos}(\theta/2) \mathrm{sin}(\theta/2) \sqrt{(1-p)(1-q)}|f|/\mathcal{P}^{ED}(t,p,q)\}.
\end{eqnarray}
By determining the strength $q$ of the post QMR,
in term of the prior WM strength $p$ and the evolution time $t$, as discussed previously,
the optimal total success probability can be rewritten as
\begin{eqnarray}
\mathcal{P}^{ED}(t,p)=(1-p)|f|^2\Bigr(1+(1-p) (1-|f|^2) \mathrm{cos}^{2}(\theta/2)\Bigr),
\end{eqnarray}
and the optimal concurrence becomes as
\begin{eqnarray}
C\bigr(\rho_{0,r}(t,p)\bigr)=2 \mathrm{max}\{0,\mathrm{cos}(\theta/2) \mathrm{sin}(\theta/2) (1-p)|f|^{2}/\mathcal{P}^{ED}(t,p)\}.
\end{eqnarray}
The time dependency of the optimal concurrence has been shown in Fig. 3 (a, c) in two situation:
in one hand, the concurrence evolves naturally, i.e. $p=q=0$, and on the other, the evolution of the concurrence takes place under the measurement-controlled dynamics.
By comparing this two situation we can observe that
the concurrence under natural evolution will eventually decay to zero,
but entanglement in our scheme can be improved considerably by controlling the WM strength $p$.
Also, it is easy to see that as the WM strengths $p$ becomes considerable so does the concurrence too.
Moreover, we observe from Fig. 3 (b, d) that the optimal success probability is decreasing in term of $p$, as we expected.

\section{Conclusion}
We have studied the QST over binary-tree spin network by applying two quantum partially collapsing measurements.
In this regard, we performed a WM with strength $p$ on the sending qubit of the network before the evolution of the system, and a QMR with strength $q$ on the receiving qubit after the evolution. By taking the optimal choice of quantity $q$ in term of $p$, it was shown that the QST can be improved considerably by controlling only the WM strength $p$, and even near-perfect QST can be achieved by choosing $p$ to be very close to 1. Finally, we have investigated the entanglement distribution quality over the network by using this approach. We found that the entanglement distribution over the binary-tree network can be obviously improved by exploiting partially collapsing measurements protocol.

\newpage
Fig. 1. Schematic representation of a binary-tree network with $N=4$ generations.
\begin{figure}
\centering
\includegraphics[width=300 pt]{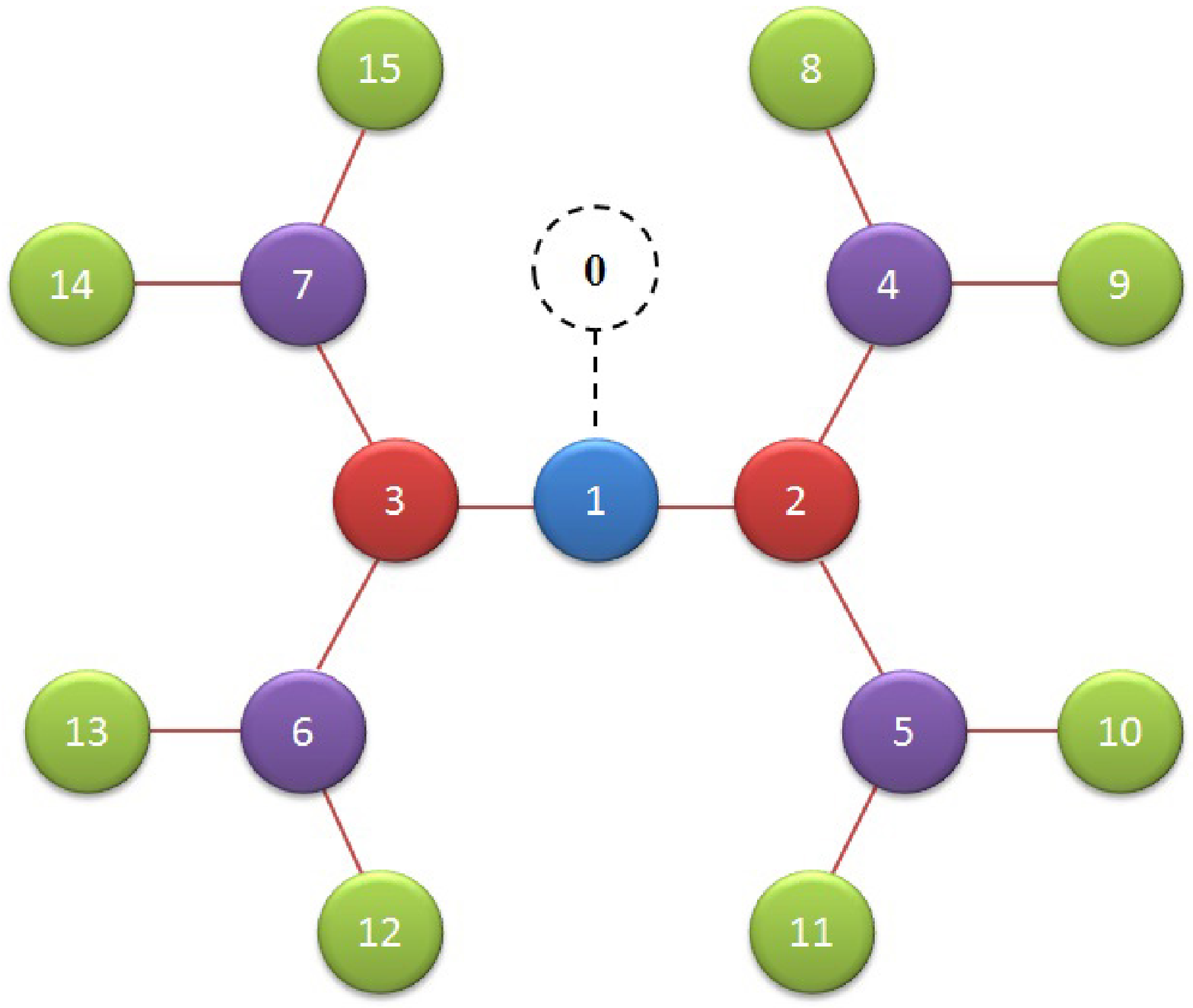}
\caption{} \label{Fig1}
\end{figure}

\newpage
Fig. 2. (a, c) Time dependency of the averaged fidelities, $F_{ave}(t,0)$ (starred  curves)
under the natural evolution and $F_{ave}(t,p)$ (dotted dashed , dashed , solid curves) under the measurement-controlled evolution with $p = 0.2, 0.6, 0.99$, respectively. (b,d) Time dependency of the averaged success probability for the different weak measurement strengths $p$. (a) and (b) panels are plotted with $N=4$ generations, (c) and (d) panels with $N=8$ generations. The parameters are $\nu=1$ and $\lambda=0.5$.

\begin{figure}
        \qquad \qquad\qquad\qquad \qquad a \qquad\qquad \quad\qquad\qquad\qquad\qquad\qquad\qquad b\\{
        \includegraphics[width=3in]{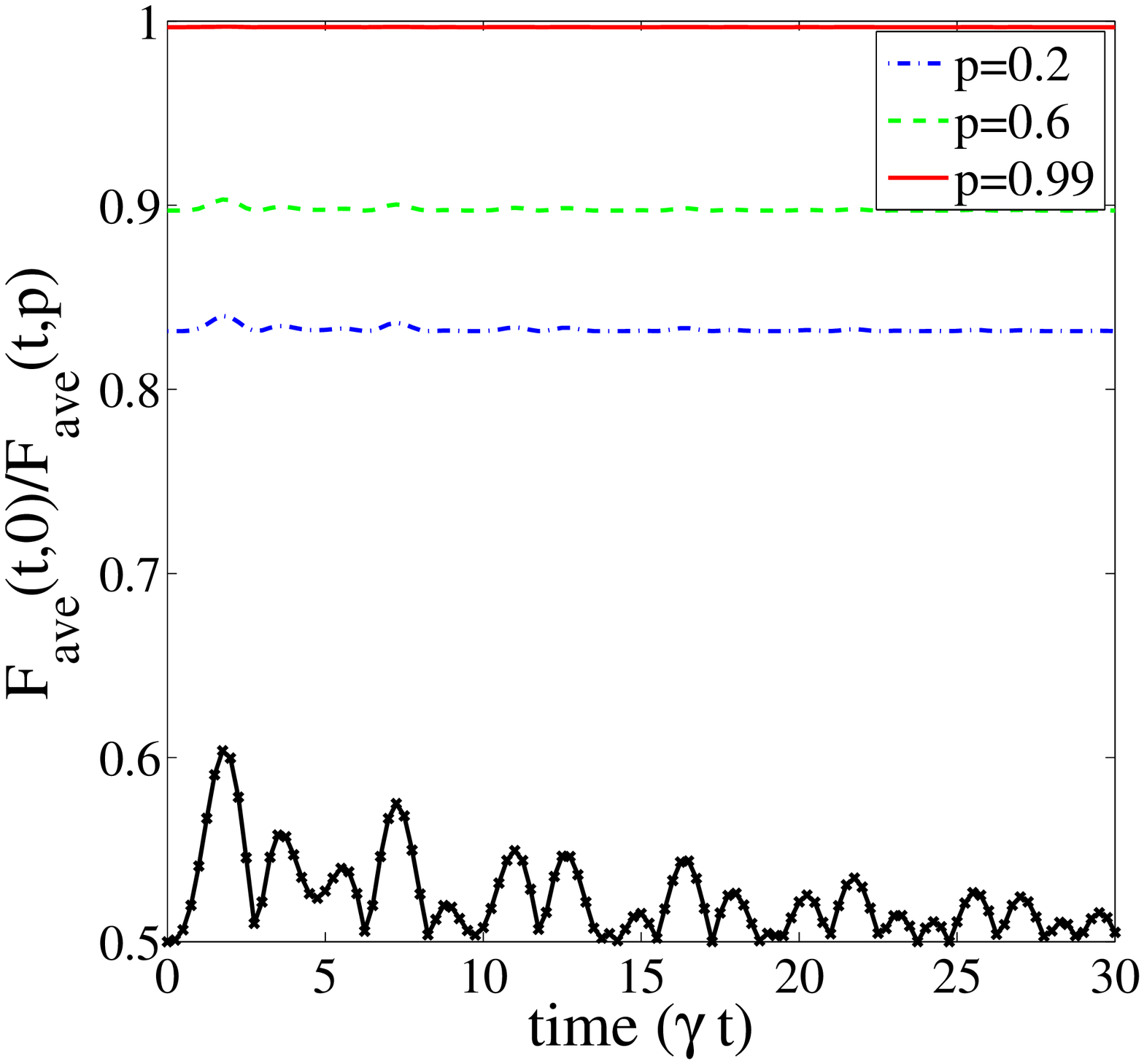}
        \label{fig:first_sub}
    }{
        \includegraphics[width=3in]{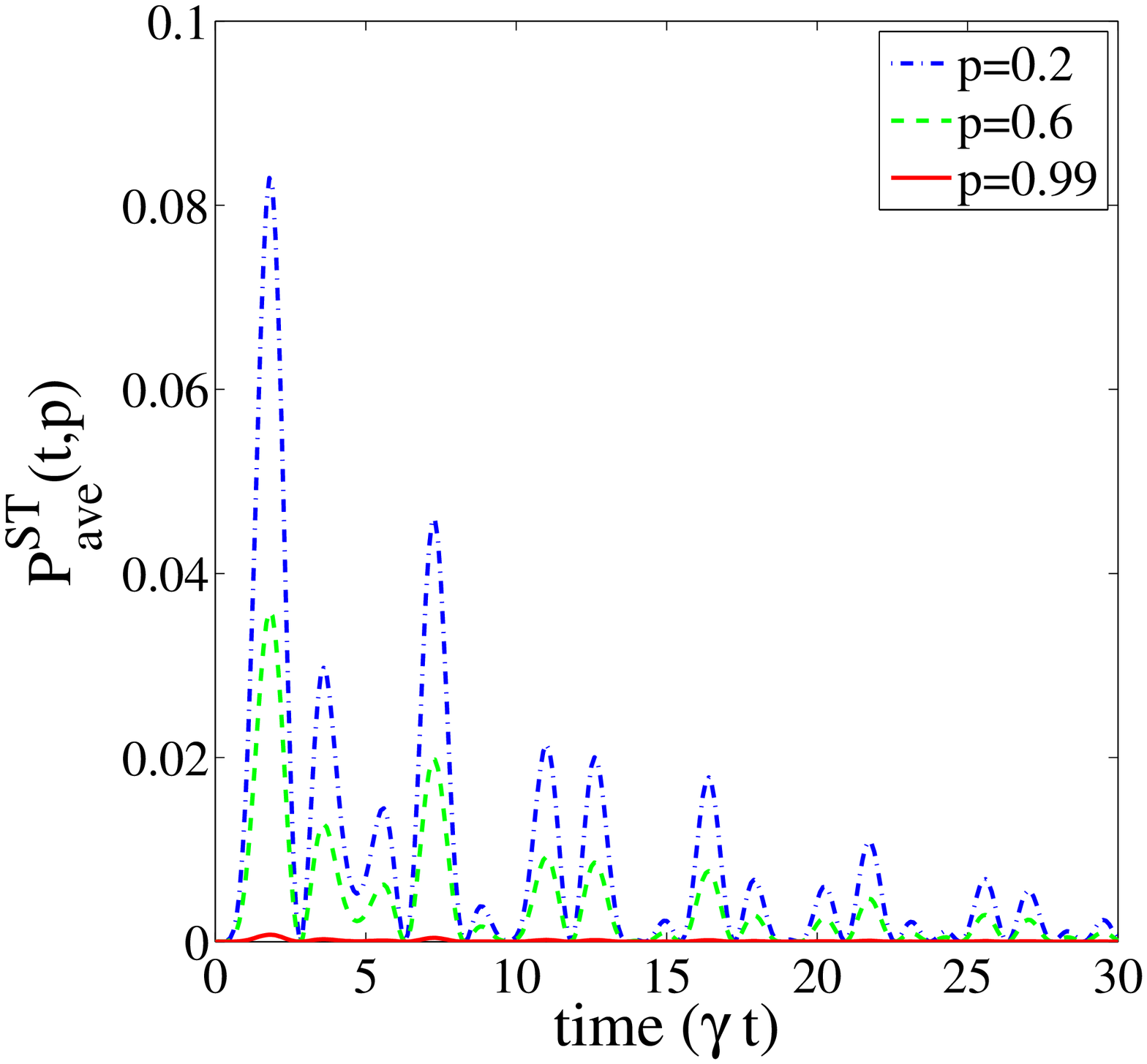}
        \label{fig:second_sub}
    }\\ \par \quad \quad\qquad\qquad\qquad \qquad c \qquad \qquad\qquad\qquad\qquad\quad\qquad\qquad\qquad d\\{
        \includegraphics[width=3in]{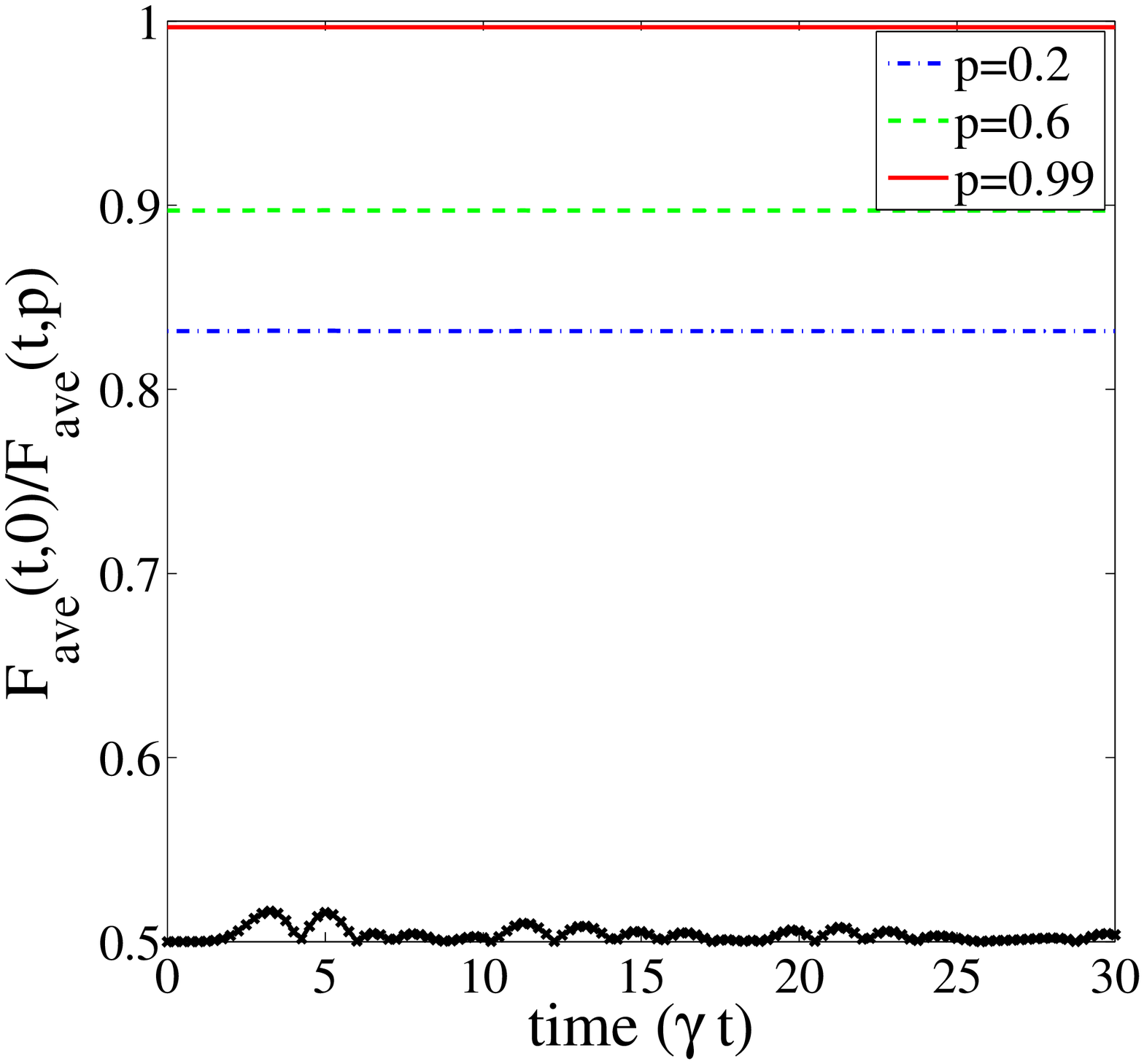}
        \label{fig:first_sub}
    }{
        \includegraphics[width=3in]{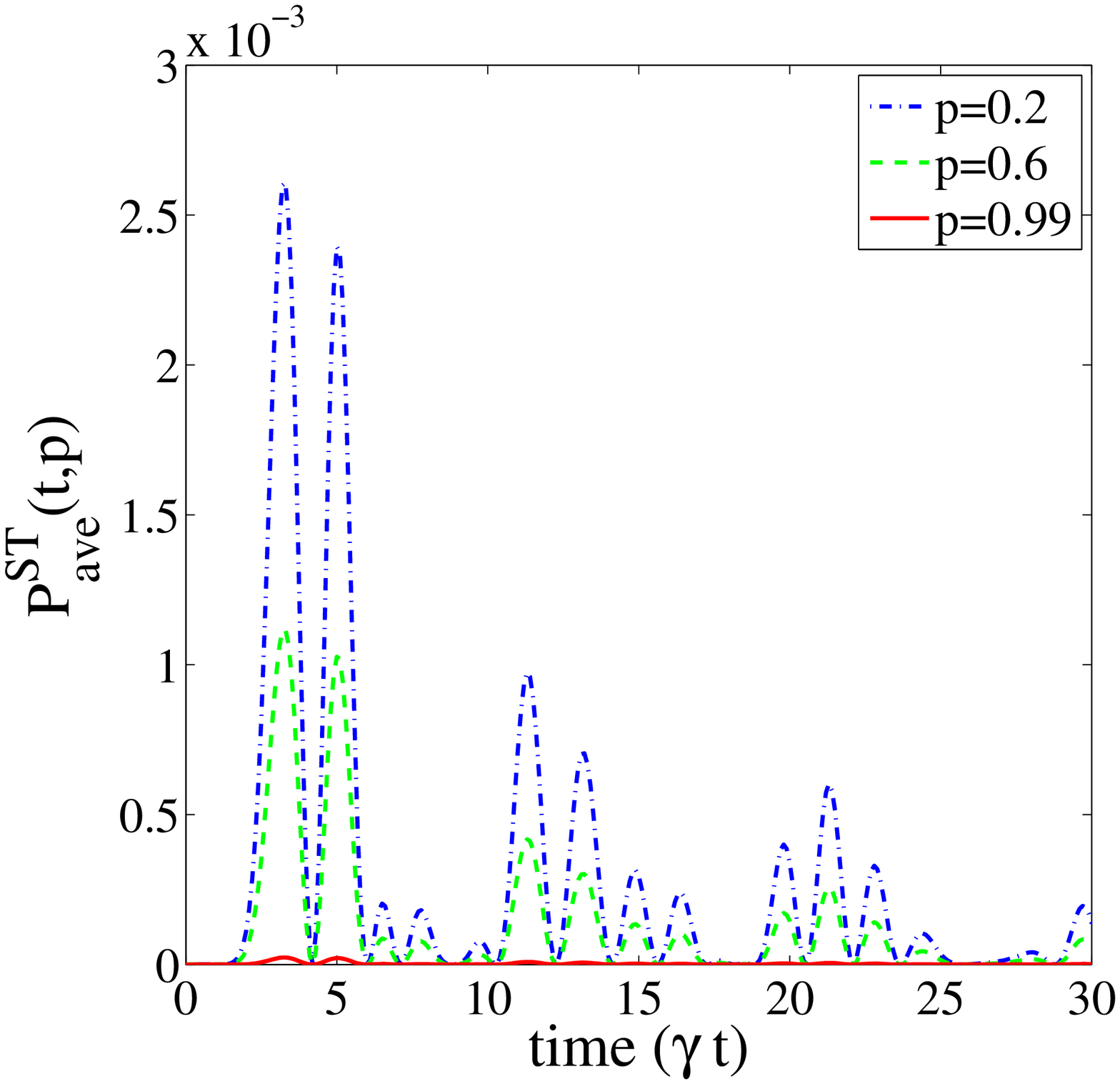}
        \label{fig:second_sub}
    }
    \caption{}
    \end{figure}

\newpage
Fig. 3. (a, c) Time dependency of the optimal concurrence, under the natural evolution (starred  curves) and
under the measurement-controlled evolution (dotted dashed , dashed ,solid curves) with $p = 0.2, 0.6, 0.99$, respectively. (b, d) Time dependency of the optimal success probability for the different weak measurement strengths $p$. (a) and (b) panels are plotted with $N=4$ generations, (c) and (d) panels with $N=8$ generations. The parameters are $\theta=\pi/2$, $\nu=1$ and $\lambda=0.5$.

\begin{figure}
        \qquad \qquad\qquad\qquad \qquad a \qquad\qquad \quad\qquad\qquad\qquad\qquad\qquad\qquad b\\{
        \includegraphics[width=3in]{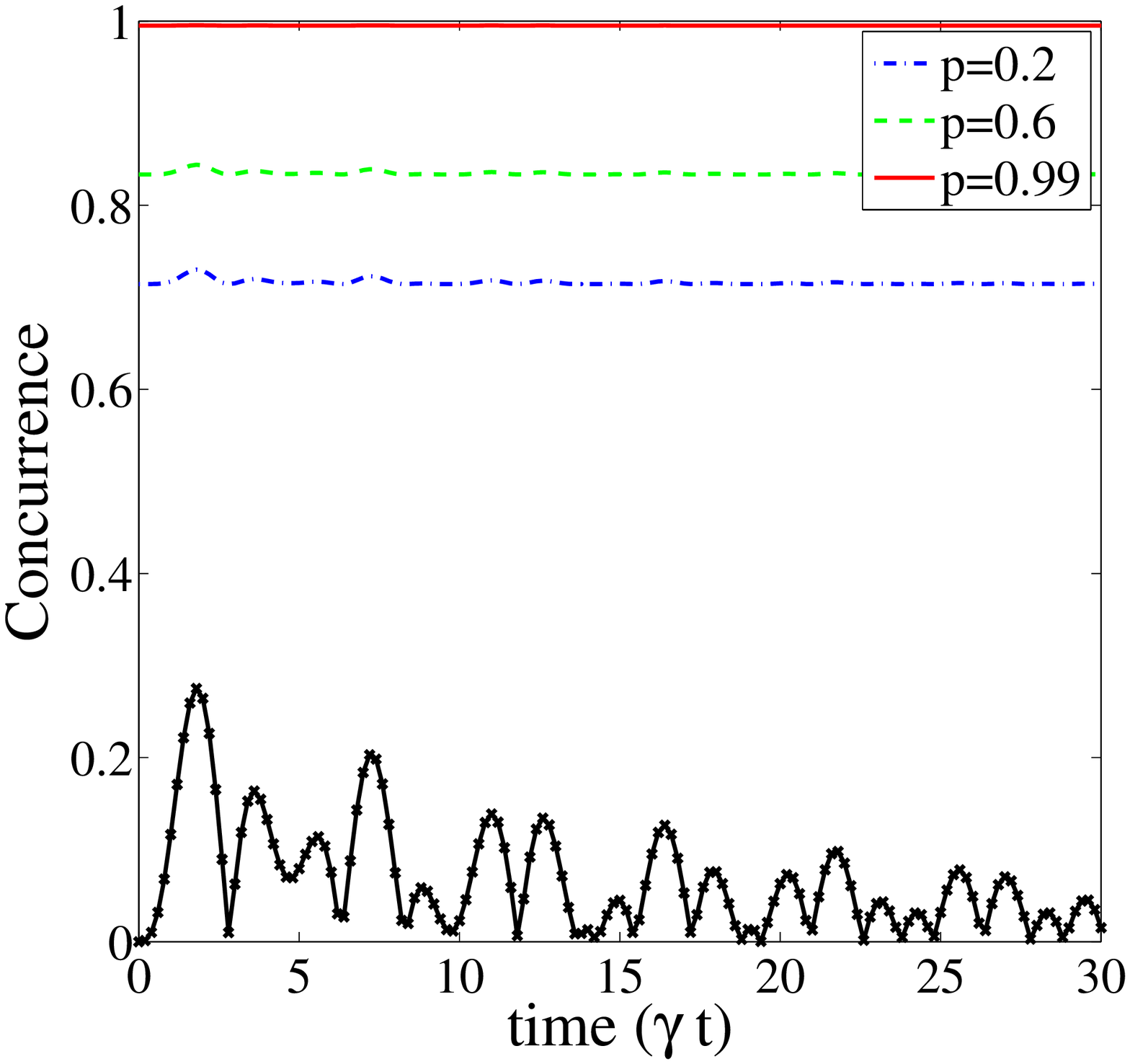}
        \label{fig:first_sub}
    }{
        \includegraphics[width=3in]{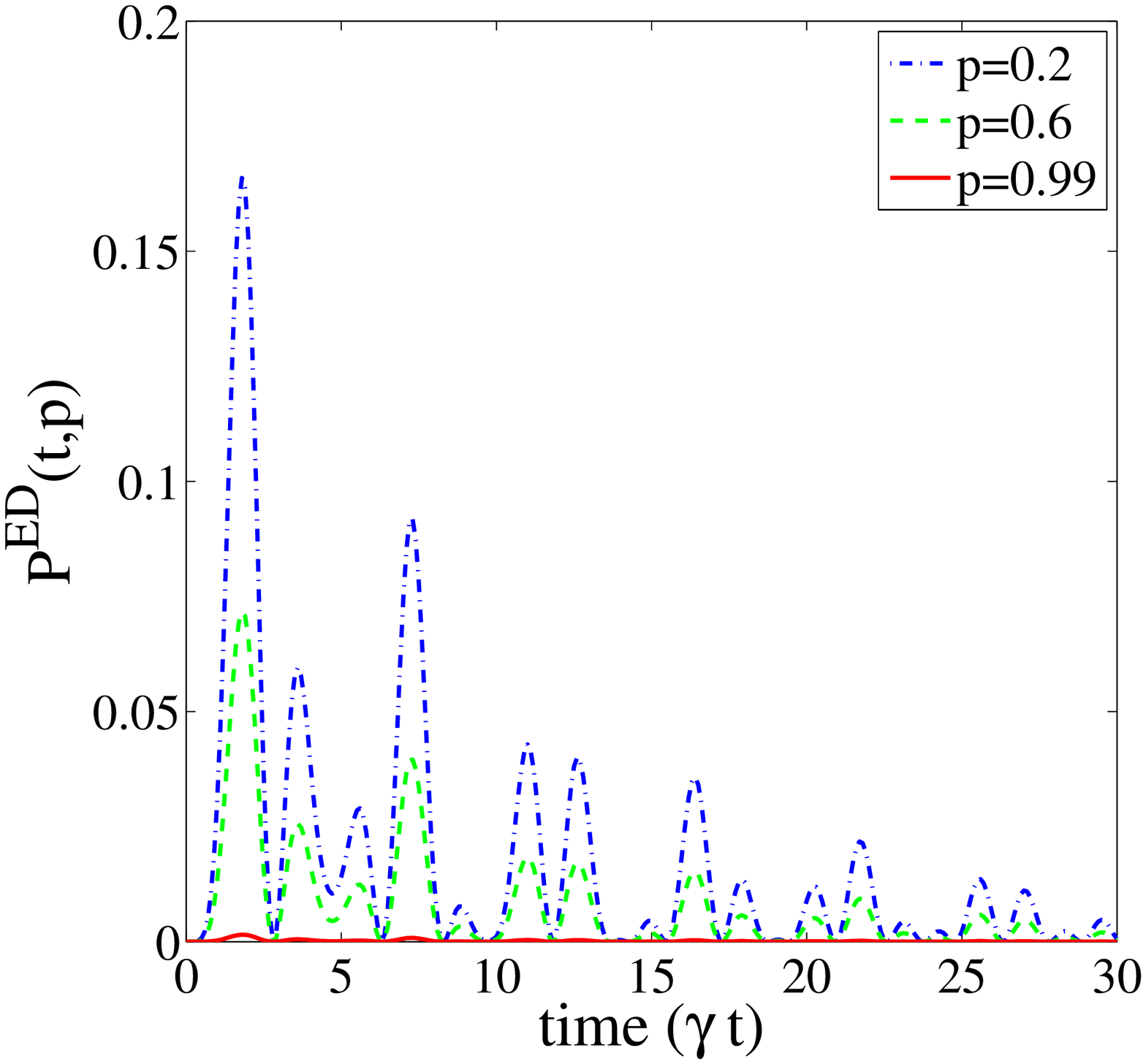}
        \label{fig:second_sub}
    }\\ \par \quad \quad\qquad\qquad\qquad \qquad c \qquad \qquad\qquad\qquad\qquad\quad\qquad\qquad\qquad d\\{
        \includegraphics[width=3in]{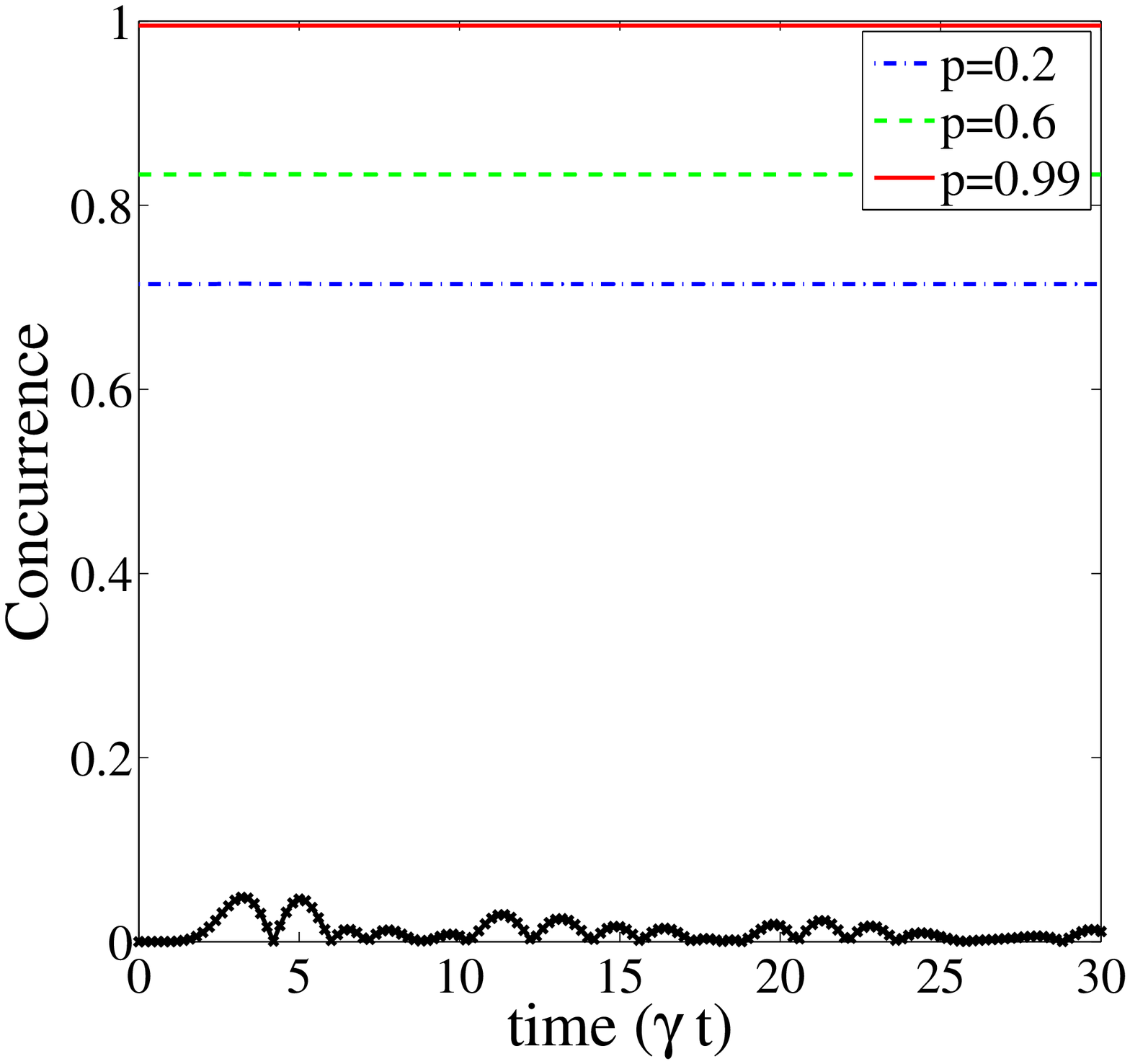}
        \label{fig:first_sub}
    }{
        \includegraphics[width=3in]{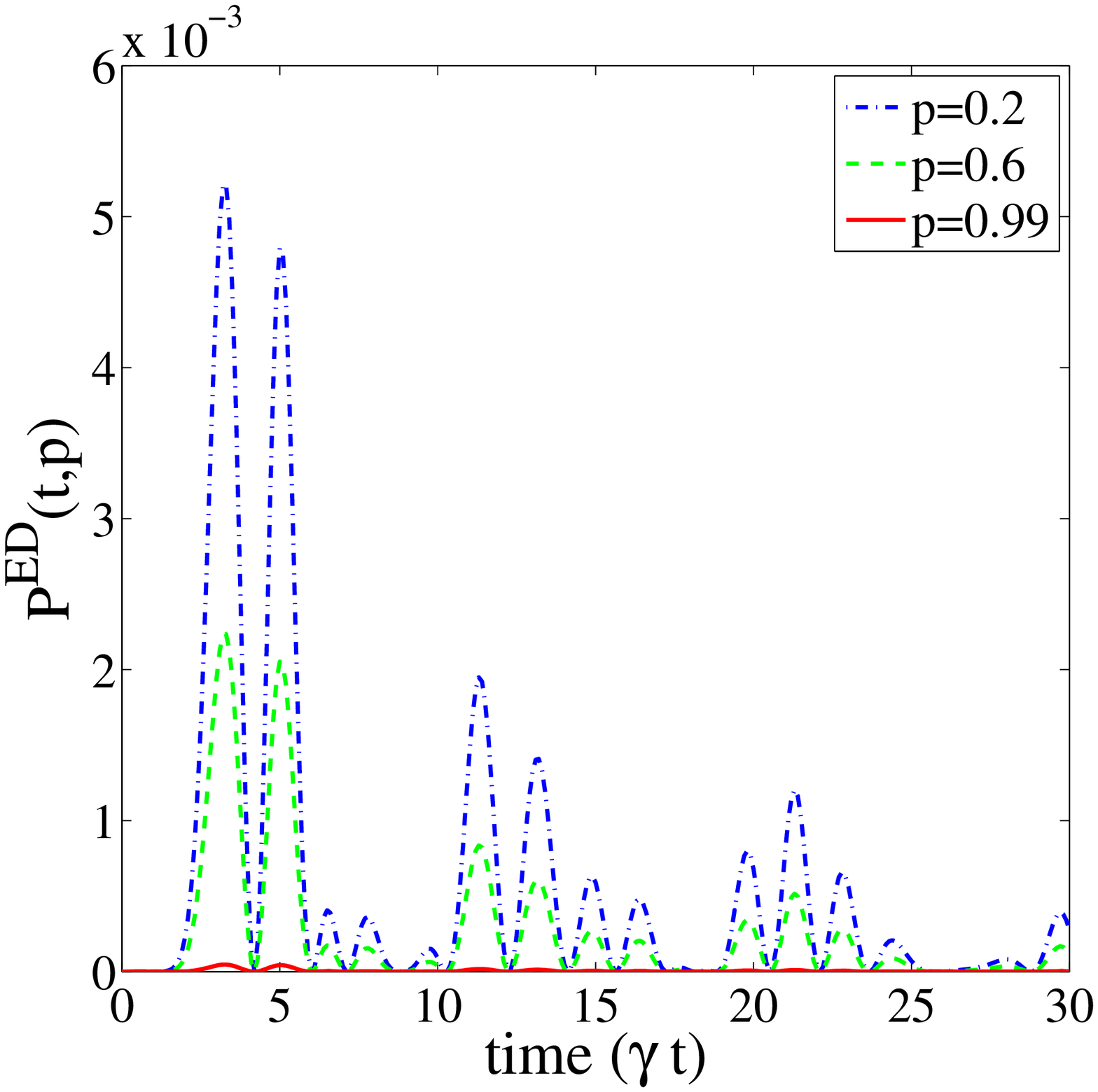}
        \label{fig:second_sub}
    }
    \caption{}
    \end{figure}

\end{document}